\documentstyle[12pt,epsfig]{article}
\newcommand\GeV{\,\mbox{GeV}}
\newcommand\beq{\begin{equation}}
\newcommand\eeq{\end{equation}}
\newcommand\bea{\begin{eqnarray}}
\newcommand\eea{\end{eqnarray}}

\newcommand\ra{\rightarrow}
\begin{document}
\begin{titlepage}

\begin{flushleft}
{\tt hep-ph/9610203} \\[0.1cm]
October   1996
\end{flushleft}
\vspace{0.2cm}
\begin{center}
\Large
{\bf  The Effect of Small-$x$ Resummations on the Evolution} \\

\vspace{3mm}
{\bf of Polarized Structure Functions} \\

\vspace{1.2cm}
\large
J. Bl\"umlein \\
\vspace{0.3cm}
\normalsize
{\it DESY--Zeuthen, Platanenallee 6, D--15735 Zeuthen, Germany} \\
\vspace{0.8cm}
\large
A. Vogt\\
\vspace{0.3cm}
\normalsize
{\it Institut f\"ur Theoretische Physik, Universit\"at W\"urzburg} \\
\vspace{0.1cm}
{\it Am Hubland, D--97074 W\"urzburg, Germany} \\
\vspace{3cm}
\large
{\bf Abstract}
\normalsize
\end{center}
\vspace{-0.3cm}
The impact of the resummation of leading small-$x$ terms in the 
anomalous dimensions is briefly summarized for the evolution of
non--singlet and singlet polarized structure functions.
\vfill 
\noindent
\small
{\sf
Contribution to the Proceedings of
the 12th Int. Symposium on High--Energy Spin
Physics,  Amsterdam, Sept. 10--14, 1996}
\normalsize
 
\end{titlepage}
%
\begin{center}
{\large \bf
 The Effect of Small-{\boldmath $x$} Resummations on the Evolution\\ 
 of Polarized Structure Functions\\
}
\vspace{5mm}
\underline{Johannes Bl\"umlein}$^1$ and Andreas Vogt$^{2}$\\
\vspace{5mm}
{\small\it
(1) DESY--Zeuthen, Platanenallee~6, D--15735 Zeuthen, Germany\\
(2) Institut f\"ur Theoretische Physik, 
Universit\"at W\"urzburg, Am Hubland, D--97074~W\"urzburg, Germany\\
 }
\end{center}

\begin{center}
\vspace{3mm}
ABSTRACT

\vspace{3mm}
\begin{minipage}{130 mm}
\small
The impact of the resummation of leading small-$x$ terms in the 
anomalous dimensions is briefly summarized for the evolution of
non--singlet and singlet polarized structure functions.
\end{minipage}
\end{center}


\vspace{3mm}
\noindent
{\large\bf 1~~Introduction}
 
\vspace{1mm}
\noindent
The evolution kernels of both non--singlet and singlet polarized parton
densities contain large logarithmic contributions for small fractional
momenta $x$. The leading terms in this limit are of the form $ \alpha _s
^k \ln^{2k-2}x $ for both cases \cite{KL,BER2}. The resummation of these
contributions to all orders in the strong coupling constant $\alpha_s$ 
can be completely derived by means of perturbative QCD. The appropriate
framework for investigating the resummation effects is provided by the 
renormalization group equations. 

The impact of the resulting all--order anomalous dimensions on the
behaviour of the deep--inelastic scattering (DIS) structure functions 
at small $x$ thus depends as well on the non--perturbative input 
parton densities at an initial scale $Q_0^2$. Hence the resummation 
effects can only be studied via the evolution over some range in $Q^2$. 
This evolution moreover probes the anomalous dimensions also at medium 
and large values of $x$ by the Mellin convolution with the parton 
densities. Therefore the small-$x$ dominance of the leading terms over 
less singular contributions in the anomalous dimensions does not 
necessarily imply the same situation for observable quantities, such 
as the structure functions.

In the following we present a brief survey of quantitative results 
which shed light on the importance of these aspects. For full accounts, 
including the discussion of theoretical aspects, the reader is referred
to refs.~\cite{BVplb1,BVcrac} and \cite{BVplb2} for the non--singlet 
and singlet evolutions, respectively. A recent review covering also the
unpolarized cases can be found in ref.\ \cite{BRVrhe}.

\vspace{4mm}
\noindent
{\large\bf 2~~Quantitative Results}
 
\vspace{1mm}
\noindent
In leading (LO) and next--to--leading order (NLO) perturbative QCD, the 
complete anomalous dimensions are known. Hence the effect of the 
all--order resummation of the most singular parts of the splitting 
functions as $x \rightarrow 0$ concerns only orders higher than 
$\alpha_{s}^2$. Due to the Mellin convolution terms less singular as 
$x \ra 0$ may contribute substantially also at these higher orders. 
Since such contributions and further corrections are not yet known to 
all orders, it is reasonable to estimate their possible impact by 
corresponding modifications of the resummed  anomalous dimensions 
$\Gamma (N,\alpha_s)$, where $N$ denotes the Mellin variable. Plausible
examples inspired by the behaviour of the full NLO results have been 
studied in refs.~\cite{BVplb1,BVcrac,BVplb2}, including
\begin{equation}
\begin{array}{clcl}
{\rm A:} & \Gamma(N, \alpha_s)
\rightarrow  \Gamma(N, \alpha_s) - \Gamma(1, \alpha_s) \:\:\:\:
 &
{\rm B:} & \Gamma(N, \alpha_s)
\rightarrow  \Gamma(N, \alpha_s)(1 - N)
\\
{\rm D:} & \Gamma(N, \alpha_s)
\rightarrow  \Gamma(N, \alpha_s)(1 - 2N + N^3)\, . & & 
\end{array}
\end{equation}
Clearly the presently known resummed terms are only sufficient for
understanding the small-$x$ evolution, if the difference of the results
obtained by these prescriptions are small.

\vspace{4mm}
\noindent
{\large\bf 2.1~~Polarized non--singlet structure functions}

\vspace{1mm}
\noindent
This case has been investigated in refs.~\cite{BVplb1,BVcrac} for the 
structure function combination $g_1^{\, ep}\! -g_1^{\, en}$ for two 
parametrizations of the non--perturbative initial distributions,
see Figure 1. 
\begin{center}
\vspace*{-19mm}
\mbox{\hspace*{-5mm}\epsfig{file=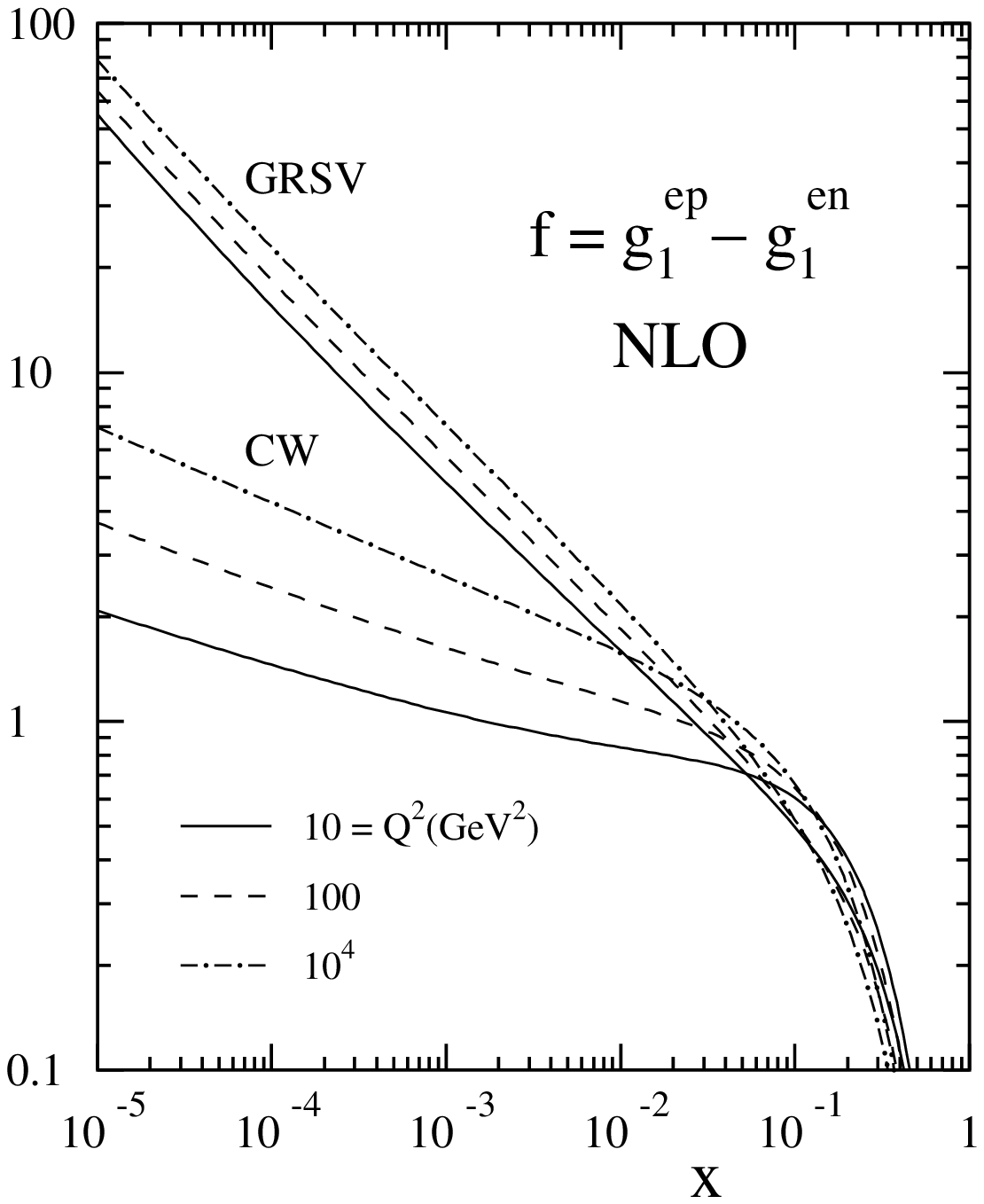,height=8.5cm,width=6.8cm}
      \hspace{1mm}\epsfig{file=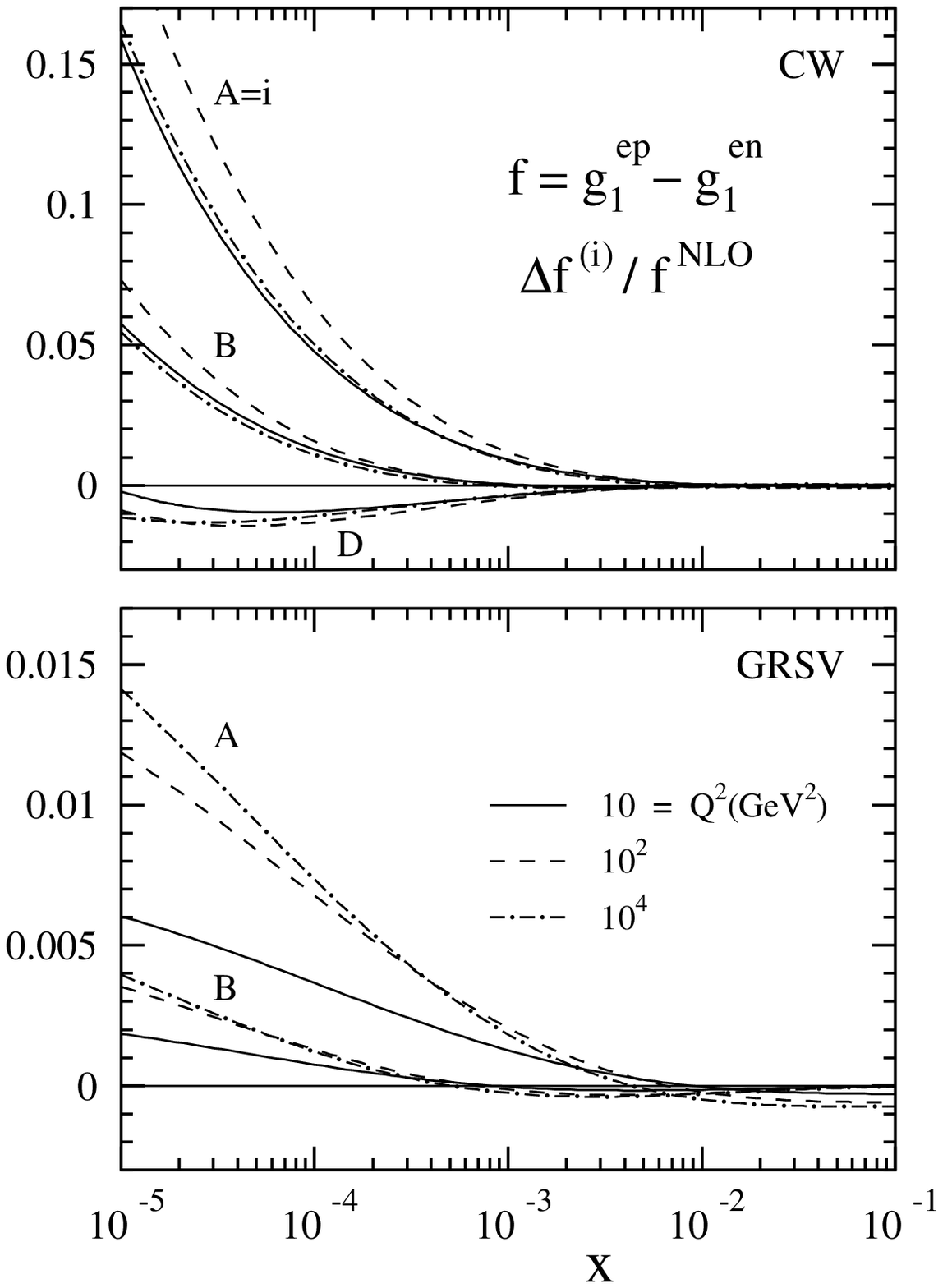,height=11.0cm,width=7.6cm}}
\vspace*{-6mm}
\end{center}
{\small
\sf Figure~1:~The NLO small-$x$ evolution of the polarized non--singlet
 structure function combination $g_1^{\, ep} - g_1^{\, en}$, and the 
 relative corrections due to the resummed kernels, for the initial
 distributions of refs.~\cite{CW} and \cite{GRSV}. The dependence on 
 possible less singular terms is illustrated by the prescriptions 
 `A', `B', and `D' of eq.~(1). The figure has been adapted from 
 ref.~\cite{BRVrhe}.}

\vspace{3mm}
\noindent
Results on the interference structure function 
$g_{5, \gamma Z}^{\, ep} (x,Q^2)$ can be found in ref.~\cite{BRVrhe}. 
For the relatively flat CW input \cite{CW}, the resummation effect on
$g_1^{\, ep} - g_1^{\, en}$ reaches about 15\% at $x = 10^{-5}$.
However, in the restricted kinematical range accessible in possible
future polarized electron--polarized proton collider experiments at
HERA \cite{JBkin}, it amounts to only  1\% or less. For the steeper
GRSV initial distributions \cite{GRSV}, the effect is of order 1\% or
smaller in the whole $x$ range. Hence the results do not at all come up 
to previous expectations of huge corrections up to factors of 10 or 
larger as anticipated in ref.~\cite{BER1}.
Note that in the $\overline{\rm MS}$ scheme the $O([\alpha_s \ln^2 x]
^l)$ terms are not present in the coefficient functions at all known 
orders~\cite{BVplb1,BVcrac}. The first moment $\Delta g_1^{\, \rm NS}
= \int_0^1 \! dx \, g_1^{\, \rm NS}(x)$, entering the Bjorken sum rule, 
depends on the coefficient functions only.

\vspace{4mm}
\noindent
{\large\bf 2.2~~Polarized singlet structure functions}

\vspace{1mm}
\noindent
The numerical consequences of the small-$x$ resummation for the 
evolution of the parton densities and $g_1^{\, ep,en}(x,Q^2)$ have been
given for different input distributions in ref.~\cite{BVplb2}. Figure~2 
shows an example. The effects are much larger here than for the 
non--singlet structure functions.
Also illustrated in these figures [$\,$by the results for the 
prescription `(B)' in eq.~(1)$\,$] is the possible impact of the yet 
uncalculated terms in the higher--order anomalous dimensions which are 
down by one power of $N$ with respect to the resummed leading pieces as 
$N \ra 0$.  As in the non--singlet case considered before, the effect 
of these additional terms can be very large; even the sign of the 
deviation from the NLO evolution cannot be taken for granted.
Moreover, for the power--law behaviour of the singlet part $g_1^{\, S}
(x)$ as $x \rightarrow 0$ estimated in refs.~\cite{BER2},
$\Delta g_1 = \int_0^1 \! dx \, g_1(x)$, which measures the charge 
weighted quark spin, is not finite. 

\begin{center}
\vspace*{-5mm}
\mbox{\epsfig{file=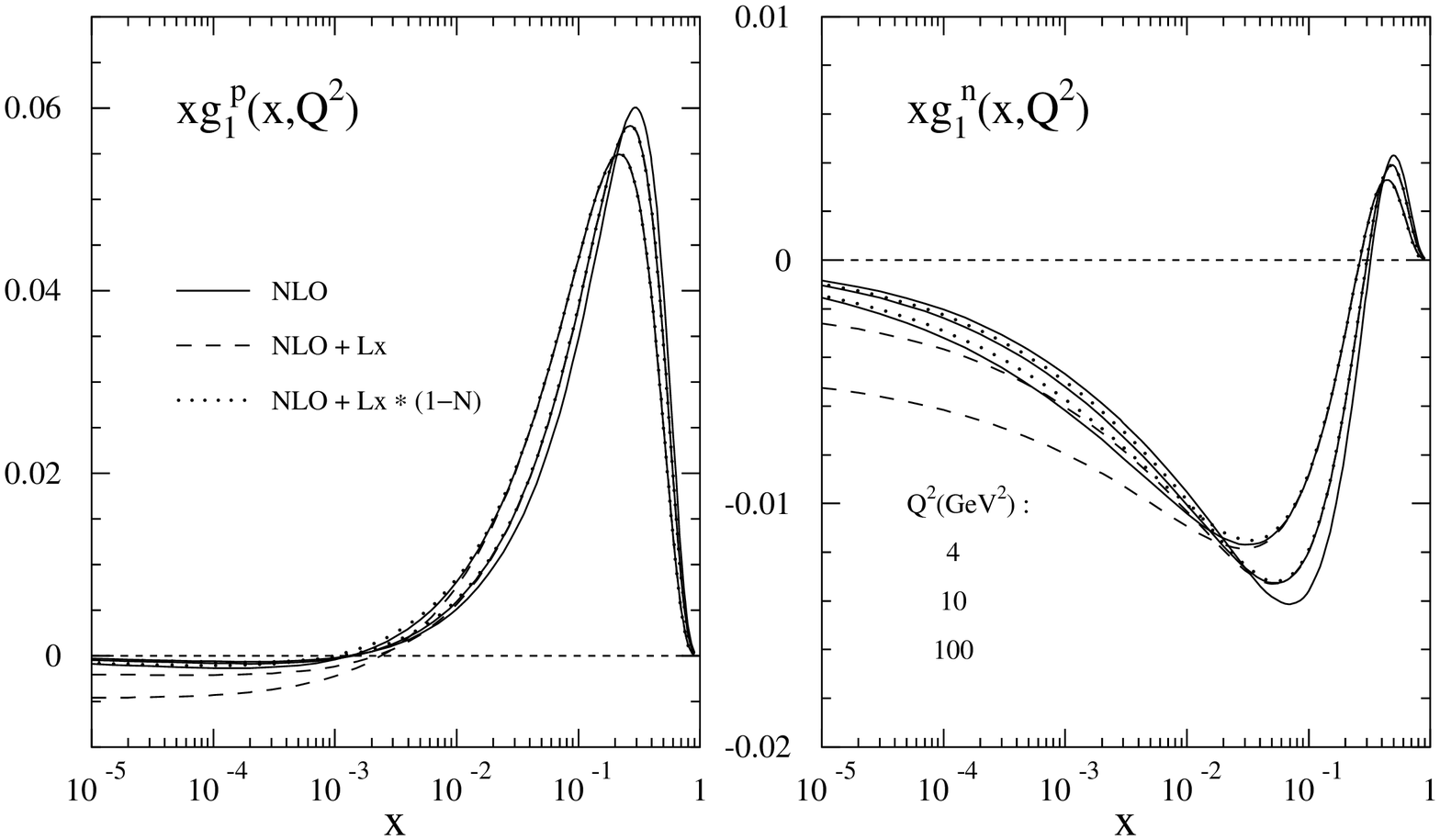,height=9.0cm}}
\vspace*{-8mm}
\end{center}
{\small
\sf Figure~2:~The $x$ and $Q^2$ behaviour of the polarized proton and
neutron structure functions $g_1^{\, p,n}(x,Q^2)$ as obtained from the 
GRSV standard distribution \cite{GRSV} at $Q_0^2= 4\GeV^2$. The results 
are shown for the NLO kernels (full), the leading small-$x$ resummed 
kernels (dashed), and the modification `B' of eq.~(1) of the latter by 
possible less singular terms (dotted). The figure has been taken from
ref.~\cite{BVplb2}.}

\vspace{2mm}
\noindent
This indicates that less singular 
terms need to contribute at the same level as the most singular ones 
even in the limit $x \rightarrow 0$.

\vspace{4mm}
\noindent
{\large\bf 3~~Conclusions}
 
\vspace{1mm}
\noindent
The effects of the resummation of the leading small-$x$ terms in the 
polarized non--singlet and singlet anomalous dimensions have been
summarized. For non--singlet structure functions the corrections due to 
those $\alpha_s(\alpha_s \ln^2x)^{l}$ contributions are about $1\%$ or 
smaller, in the kinematical ranges probed so far as well as the regime 
accessible at a polarized version of HERA \cite{BVplb1,BVcrac}.
In the singlet case very large corrections are obtained. As in the 
non--singlet cases, however, possible less singular terms in higher 
order anomalous dimensions, are hardly suppressed against the presently 
resummed leading terms in the evolution: even a full compensation of 
the resummation effects cannot be excluded \cite{BVplb2}.
To draw firm conclusions on the small-$x$ evolution of also the singlet 
structure functions, the next less singular terms as well as the 
complete three--loop anomalous dimensions are needed.
\vspace{3mm}

\noindent {\bf Acknowledgements :} This work was supported in part
by the EC Network `Human Capital and Mobility' under contract No.\
CHRX--CT923--0004 and by the German Federal Ministry for Research and
Technology (BMBF) under contract No.\ 05 7WZ91P (0).

\small

\end{document}